\documentclass[aps,prb,twocolumn,showpacs,10pt]{revtex4-1}
\usepackage{graphicx,amsmath,amssymb,amsfonts,sidecap,color}

\makeatletter
\newcommand*{\rom}[1]{\expandafter\@slowromancap\romannumeral #1@}
\makeatother

\begin{document}

\title{Integer and Fractional Quantum Hall Effect in a Strip of Stripes}

\author{Jelena Klinovaja}
\affiliation{Department of Physics, University of Basel,
             Klingelbergstrasse 82, CH-4056 Basel, Switzerland}
\author{Daniel Loss}
\affiliation{Department of Physics, University of Basel,
             Klingelbergstrasse 82, CH-4056 Basel, Switzerland}

\date{\today}

\begin{abstract}
We study anisotropic stripe models of interacting electrons in the presence of magnetic fields in the quantum Hall regime with integer and fractional filling factors. The model 
consists of an infinite strip of finite width that contains periodically arranged stripes (forming supercells) to which the electrons are confined and between
 which they can hop with associated magnetic phases. The interacting electron system within the one-dimensional stripes are described by Luttinger liquids and shown to give rise to charge and spin density waves that
lead to periodic structures within the stripe with a reciprocal wavevector $8k_F$ in a mean field approximation.
This wavevector gives rise to Umklapp scattering and resonant scattering that results in gaps
and chiral edge states at all known integer and fractional filling factors $\nu$.
The integer and odd denominator filling factors arise  for a uniform distribution of stripes,
whereas the even denominator filling factors arise for a non-uniform stripe distribution. 
We focus on the ground state of the system, and identify the quantum Hall regime via the quantized Hall conductance.
For this we calculate the Hall conductance via the Streda formula and show that it is given by $\sigma_H=\nu e^2/h$ for all filling factors. 
In addition, we show that the
composite fermion picture follows directly from the condition of the resonant Umklapp scattering. 
\end{abstract}

\maketitle

\section{ Introduction}  Topological effects in condensed matter systems  were extensively studied over the last decades with particular focus on the physics of  quantum Hall systems  and, more recently, on topological insulators \cite{Wilczek,Hasan_RMP,Zhang_RMP,Alicea_2012}. A hallmark of these systems is the robust quantization of the Hall conductance, as observed first in a two-dimensional electron gas (2DEG) in the presence of a perpendicular magnetic field \cite{Klitzing,Tsui_82}  giving rise to the integer and fractional quantum Hall effect (IQHE and FQHE) \cite{QHE_Review_Prange,book_Jain}. A further hallmark of topological systems are edge states that can give rise to exotic quantum states with non-Abelian statistics such as Majorana fermions \cite{Wilczek,Hasan_RMP,Zhang_RMP,Alicea_2012,Read_2000,Nayak,fu,Nagaosa_2009,Sato,lutchyn_majorana_wire_2010,oreg_majorana_wire_2010,Klinovaja_CNT,Sticlet_2012,bilayer_MF_2012,MF_nanoribbon}. 
While the IQHE arises for non-interacting electrons \cite{QHE_Review_Prange,book_Jain,Laughlin,Thouless,Halperin} and can be characterized by topological Chern numbers on the lattice, \cite{Thouless} the FQHE is a property of interacting electrons in  fractionally filled Landau levels \cite{QHE_Review_Prange,book_Jain,Laughlin_FQHI}.
In an attempt to extend these concepts beyond the traditional systems such as 2DEGs, 
a remarkable series of new models with similar topological properties have been proposed in recent years,  
such as FQHE-like systems and topological Chern insulators, many of which have the potential to be implemented  with cold atoms and molecules, where the magnetic field effects can be
efficiently mimicked by a high control of phases. 
\cite{Demler_Lukin_2005,Demler_Lukin_2007,Rudner_Demler_2010,Dalibard_RMP_2011,Chamon_Mudry_2011,Nagaosa_Okamoto_2011,Daghofer_2012,Cooper_Dalibard_2013,Yao_Lukin_2013,oded_2013,QHE_Klinovaja_Loss,kane_PRL,Kane_lines}

Moreover, the Hall effect was also predicted and  observed in quasi-two-dimensional materials such as organic compounds, for example, Bechgaard salts \cite{Lebed,Montambaux,Lebed_Gorkov,Yakovenko_PRB,Lee_PRB,Yakovenko_review,Stripe_PRL_exp}. The advantage of such structures lies in their highly anisotropic behaviour which arises due to the stacking of quasi-one-dimensional molecules in the plane such that the hopping matrix element along the molecule direction is much  larger than the one in the perpendicular direction (and negligible hopping between the planes), which can be modeled
as two-dimensional  lattice with anistropic spectrum. \cite{Yakovenko_review}
As a result, the integer quantum Hall effect observed in these materials cannot be understood in terms of the isotropic Landau level theory developed for semiconducting two-dimensional electron gases and requires a separate treatment \cite{Yakovenko_review}. Importantly, such a strong anisotropy,  which seems to be challenging to be treated analytically at first sight, opens up a new platform for the theoretical treatment 
of both integer and fractional quantum Hall effect. While the IQHE has been understood  in  such  anistropic systems long ago, in particular in pioneering work by Yakovenko, \cite{Yakovenko_PRB,Yakovenko_review} the FQHE  has been addressed only recently by
Kane {\it et al.}, \cite{kane_PRL,Kane_lines} who included for the first time electron-electron interactions and showed that the anisotropic system, modeled as tunnel-coupled wires, can support fractional quantum Hall  states, including not only the Haldane-Halperin hierachy (observed in two-dimensional electron gases), but also additional ones. \cite{kane_PRL,Kane_lines,Kane_lines_footnote}
Such anisotropic models have the advantage that they allow one to treat the interactions in one direction exactly by making use of Luttinger liquid formalism. \cite{kane_PRL,Kane_lines,Yaroslav}

  All this taken together, namely the availability of highly anisotropic materials and of a powerful theoretical apparatus to treat interactions, serves as the main motivation to turn our attention to quantum Hall effect in an anisotropic strip of stripes model. We will show that we can obtain a rather simple and unifying
 picture of both, the integer and fractional QHE. Including strong interaction effects, we will find the fractional filling states of the Haldane-Halperin hierarchy (and only those).
 While the results obtained here for the IQHE are implicitly contained in earlier work, \cite{Yakovenko_PRB,Yakovenko_review} our approach has a straighforward extension from the IQHE regime
 to the FQHE regime by allowing for single-particle Umklapp scattering produced by strong electron-electron interactions in a mean field approxiamtion.

In line with these developments, we have recently discovered \cite{QHE_Klinovaja_Loss}  a new connection between edge states in quantum Hall systems  and fractionally charged fermions \cite{Jackiw_Rebbi,FracCharge_Su,FracCharge_Kivelson,CDW,Two_field_Klinovaja_Stano_Loss_2012,Klinovaja_Loss_FF_1D} and Majorana fermions \cite{Read_2000,Nayak,fu,Nagaosa_2009,Sato,lutchyn_majorana_wire_2010,oreg_majorana_wire_2010,Klinovaja_CNT,Sticlet_2012,bilayer_MF_2012,MF_nanoribbon} that emerges upon dimensional reduction from two to one dimensions.  
The models proposed consist of highly anisotropic lattices in the presence of a magnetic flux. \cite{QHE_Klinovaja_Loss}
We showed that the QHE can be described in terms of  resonant  scattering inducing Peierls gaps, where Umklapp scattering involving higher Brillouin zones plays a crucial role, especially for the FQHE.
 In particular, we demonstrated that if the lattice is at quarter filling, 
 which was a special assumption,  gaps in the spectrum are opened at the filling factors $\nu=n/m$ ($n,m$ integers and $m$ odd), with associated chiral edge states within the gap,  in close analogy to topological band insulators. Thus, quite remarkably, this simple model reproduces all odd denominator FQHE filling factors $\nu$. However, due to the fixed lattice periods, the Hall conductance is a periodic function of the magnetic field $B$ \cite{Hofstadter}, and thus the Hall conductance can assume only integer values expressed as Chern numbers \cite{Thouless}. However, when allowing for interactions, and again assuming quarter filling for the lattice, the periodicity becomes dependent on the Fermi wavevector $k_F$. As a consequence, the associated Hall conductances given by the fractional values, $\sigma_H=\nu e^2/h$, emerge \cite{QHE_Klinovaja_Loss}, corresponding to plateaus on the classical curve $\sigma_H \propto 1/B$,
exactly as observed in 2DEGs \cite{Klitzing,Tsui_82}.  We note that in the present work we focus on the ground state of the stripe model and characterize its quantum Hall effect via the quantized Hall conductance. The nature and properties of the excitations in the stripe model  such as fractional charge, braiding statistics, relation to the Laughlin states, etc. is an interesting question by itself, which, however, we shall not address here but leave for later work. Similarly, we do not consider disorder effects and assume ballistic systems.
 
 The aforementioned result, however, depends crucially on the
 basic assumption of a quarter filled lattice, which is seen to be equivalent to a special value of one of the lattice constants, $a_x=2\pi/8k_F$. Different lattice fillings give rise to different hierarchies of FQHE series.
 Here, by going a decisive step beyond our previous work, we show that this special property could emerge naturally due to  interactions in a stripe model. The model consists of an infinite strip of 2DEG that contains a periodic arrangement of quasi one-dimensional stripes, see Fig. \ref{fig:tb_model}. 
By modeling the stripes in the framework of  Luttinger liquids, we are able to argue in terms of a mean-field approximation that charge (CDW) or spin (SDW) density waves indeed might favour the special period $a_x=2\pi/8k_F$.  
Again, exactly this special reciprocal wavevector $8k_F$ results in  Hall plateaus on the classical curve $\sigma_H \propto 1/B$, not only for the  odd denominator but also  for all even denominator FQHE states, as we will show here.

The anisotropic stripe model considered here allows us to treat the hopping between the stripes perturbatively, while the interacting electron gas within the quasi one-dimensional stripe can be described
non-perturbatively by a Luttinger liquid approach \cite{Kane_lines_footnote}.
The overall guiding principle in analyzing the stripe model is to look for the requirements that minimize the energy of the system and lead to opening of band gaps (with chiral edge states). As we shall see, this amounts to
 find those periodic arrangements of stripes in form of supercells that will allow resonant scattering within a given stripe.  The requirement for such resonances is intimately connected
 to  Umklapp scattering and to the magnetic phases (and thus to the filling factor) associated with the hopping between the stripes. The stripe arrangements for the IQHE and for odd
 denomonator FQHE are uniform, whereas they are non-uniform for even denominator FQHE. 
 It is quite remarkable that such a simple and transparent requirement of resonant scattering is all that is needed  to obtain the IQHE as well as the FQHE for even and odd denominators
 without further assumptions.

Our results can be viewed in two ways. On one hand, we demonstrate that our stripe model exhibits the IQHE and FQHE with gaps and Hall conductances at fractional fillings factors. 
We also show that the composite fermion sequences immediately follow from the resonant scattering condition.
All this by itself is a remarkable property of a simple model. 
On the other hand, the stripe model might provide a unifying picture of the Hall plateaus observed in 2DEGs under the assumption that 2DEGs form such a periodic pattern of stripes in order to minimize the potential energy.
We do not attempt to give a proof of this latter assumption, but would like to point out that such periodic structures 
can be expected to be increasingly favoured by increasing the magnetic field. Indeed,  
at high magnetic fields  interaction effects get strongly enhanced and 
electrons tend to order themselves into periodic structures. 
\cite{book_Jain,Wigner_Girvin,Wigner_Kivelson,Halperin_crystall,anisotropy_CDW,Wigner_Jain}
This expectation is also supported by the recent observation \cite{anisotropy_West,QHE_strips} of similar anisotropies in 2DEGs already at much lower magnetic fields, predicted to result from charge density waves in high Landau levels $\nu \gg 1$ \cite{anisotropy_CDW,strips_Review}.
 Moreover, we have checked our results numerically also in the isotropic limit, where hopping in $x$ and $y$ directions are the same, and see that 
the bulk gap does not close as long as the resonant scattering condition is satisfied, additionally confirming the topological stability of our results, in particular
of the edge states.
 It would be interesting to test the density profiles in 2DEGs experimentally, for instance in graphene where
 the surface can be accessed directly \cite{Amir_graphene}.

Let us now briefly describe the model and the main results derived in the following sections.
As stated before we consider an infinite anisotropic strip in the quantum Hall regime. The electrons organize themselves into stripes that are initially uniformly distributed (single period) and twofold degenerate in spin. The  resonant scattering between left and right movers, where the momentum difference is supplied by the magnetic field, leads to the opening of  gaps at particular values of the magnetic field corresponding to odd denominator FQHE filling factors. 

In the regime of the IQHE at filling factor $\nu=2n$, where $n$ is an integer, the system is spin unpolarized. In the regime of IQHE at filling factor $\nu=2n+1$ the system restructures itself to become spin polarized. Here, by spin polarized we mean any spin order, as long it is uniform along the $y$ direction (so that spin is conserved in the hopping process). This includes both a fully spin polarized state and a spin density wave (SDW). Such a SDW can be created by electron-electron interactions within a stripe. The  rotation period $K$ creates an effective periodic potential within a stripe in the $x$ direction. Similarly, a charge density wave (CDW)  results in a periodic potential along the stripe. What is most important here is that the  period of such a potential depends directly on the Fermi wavevectror, 
$K=8k_F$. The presence of a periodic potential widens the range of allowed resonant magnetic fields. In particular, the FQHE filling factors $\nu=n/m$, with $m$ being an odd integer, will emerge. Similar to the IQHE, if $n$ is an odd integer, the system is spin polarized, and if $n$ is an even integer, the system is spin unpolarized. However, if the Zeeman energy becomes relevant, the system can undergo a transition into a spin polarized state. In all case considered, we show via the Streda formula that the Hall conductance  is given by $\sigma_H = \nu e^2/h$. Moreover, in each gap there are chiral edge states propagating along a given edge in a direction determined by the magnetic field and the filling factor $\nu$.

The FQHE at even denominator filling factors is more subtle as it requires a reconstruction of stripe arrangements. In particular, the stripes are distributed non-uniformly in the supecell. As a result, the possibility to observe such fractions crucially depends on system parameters and absence of disorder.

The paper is organized as follows. In Sec. \ref{sec:model}, we introduce the stripe model. In Sec. \ref{sec:hamil}, we derive the effective linearized Hamiltonian. The resonant scattering leads to an opening of a gap with edge states inside it. Next, we address the IQHE in Sec. \ref{IQHE} and the FQHE with odd denominators in Sec. \ref{sec:order}. In these two sections we explore the resonant scattering conditions and determine whether the system is spin polarized or not. Moreover, we find quantized values of the Hall conductance using the Streda formula. Finally, in Sec. \ref{sec:even} the even denominator FQHE is discussed.  We present our conclusions in Sec. \ref{sec:conc}.

\section{Strip of Stripes \label{sec:model}} 
\begin{figure}[!tb]
 \includegraphics[width=\columnwidth]{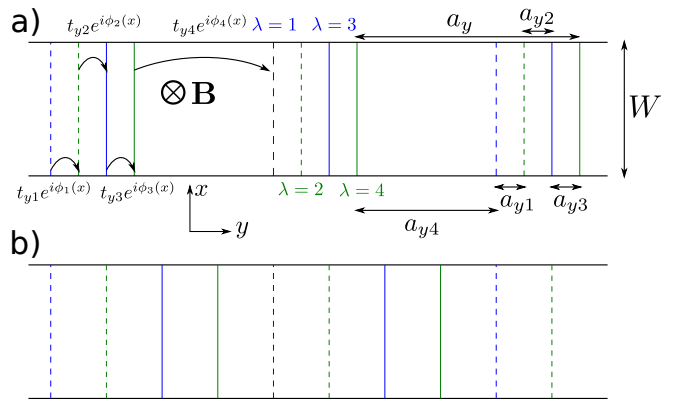}\\
 \caption{Sketch of the stripe model consisting of a infinite strip of  width $W$ with  quasi-one-dimensional stripes periodically arranged along the $y$ direction in form of supercells.
  The supercells form a one-dimensional lattice in the $y$ direction  with lattice constant $a_y$, and each supercell consists of $\Lambda$ sites 
  labeled by $\lambda=1,2,...,\Lambda$ (here $\Lambda=4$).
  The  electrons within a stripe form a continuum in $x$ direction, while the stripes are weakly tunnel connected along the $y$ direction with hopping amplitudes 
  $t_{y\lambda} e^{i\phi_\lambda(x)}$ that carry the magnetic phase $\phi_\lambda(x)$, arising from a perpendicular magnetic field
${\bf B}$.  
(a) The `dimple' or supercell formation at the filling factor $\nu=1/2$. Four sites in the supercell are non-uniformly distributed: $a_{y1}=a_{y2}=a_{y3}=a_{y4}/5=a_y/8$. 
(b) At the filling factor $\nu=1$ the stripe distribution is uniform with $a_{y\lambda}=a_y/4$.} 
 \label{fig:tb_model}
\end{figure}

We consider itinerant electrons confined to  an infinite strip in the $xy$-plane in the presence of a magnetic field $\bf B$ applied along the $z$ direction, see Fig.~\ref{fig:tb_model}. 
The strip has a finite width $W$ in the $x$ direction and is extended along the $y$ direction.  In addition, the electrons are confined into stripes, aligned parallel to the $x$ axis, and move freely inside them (no externally applied periodic potential, but see below). The stripes are weakly tunnel coupled to each other. This results in an effective  tight-binding model along the $y$ direction  that describes hopping between neighbouring stripes, while each stripe is described as a continuum in the $x$ direction. Next, we assume that $\Lambda$ stripes separated by a distance $a_{y \lambda}$ form a supercell that is periodically repeated along $y$ with period (lattice constant) $a_y=\sum_{\lambda=1}^\Lambda a_{y\lambda}$, see Fig. \ref{fig:tb_model}.

Thus, every stripe is labeled by three indices $(n,\lambda,s)$, where $n$ denotes the position of the supercell along the $y$ axis, $\lambda$ the position inside the supercell, and $s=\pm 1$ denotes the value of the spin projection in the $z$ direction.

The free-particle spectrum inside a stripe $(n,\lambda,s)$  is given by $\epsilon(k_x) = \hbar^2 k_x^2/2 m_e + \mu$, where  $k_x$  is the momentum, $m_e$  the electron mass, and $\mu$  the chemical potential. In this work we ignore the constant shift in energy due to the Zeeman term.~\cite{footnote_1} The corresponding Hamiltonian is written in standard second quantization
notation as
\begin{equation}
H_x =\int dx \sum_{n, s; \lambda=1}^\Lambda\Psi^\dagger_{n,  \lambda, s}(x) \Big[-\frac{\hbar^2 \partial_x^2}{2m_e} + \mu\Big]  \Psi_{n, \lambda, s}(x),
\label{h_x}
\end{equation}
where $\Psi_{n,\lambda,s}^\dagger(x) $  is the creation operator creating an electron with spin $s$ at the position $x$ in the stripe $(n,\lambda)$.

The hopping in the $y$ direction with amplitude $t_{y\lambda}$ (assumed to be real and positive)  is described by 
\begin{align}
&H_{y} = \int dx \Big[\sum_{n,s; \lambda=1}^{\Lambda-1} t_{y\lambda} e^{i\phi_\lambda(x)}\Psi^\dagger_{n,\lambda+1,s} (x)  \Psi_{n, \lambda,s} (x) \nonumber\\
&\hspace{20pt}+\sum_{n,s} t_{y\Lambda} e^{i\phi_\Lambda(x)}\Psi^\dagger_{n+1,  1,s}(x)  \Psi_{n, \Lambda,s}(x)+ H.c.\Big].
\label{1d_y}
\end{align} 
The phase $\phi_\lambda (x)$ is generated by the uniform magnetic field  ${\bf B}=\nabla \times {\bf A}$, where  ${\bf A}=(0,B x,0)$ is the corresponding electromagnetic vector potential
in the Landau gauge. From this we obtain the phase $\phi_\lambda (x)=(e/\hbar c)\int d{\bf r}\cdot {\bf A}$ where the integration path is a straight line in the $y$ direction at position $x$ connecting two neighboring stripes. This gives 
\begin{equation}
\phi_\lambda (x) = \frac{e}{\hbar c} B x a_{y\lambda}.
\label{phase_define} 
\end{equation}

The strip is translationally invariant in the $y$ direction. Thus, we can introduce the momentum $k_y$ via Fourier transformation, 
\begin{equation}
 \Psi_{n,\lambda,s}(x)  = \frac{1}{\sqrt{N_y}}\sum_{k_y} e^{i n k_y a_y} \Psi_{k_y, \lambda,s}(x),
\end{equation}
where $N_y$ is the number of  lattice sites in the $y$ direction.
The Hamiltonian ${H}_{y}$ is diagonal in $k_y$-momentum space, i.e., $H_{y} = \sum_{k_y} H_{y,k_y}$, where
\begin{align}
&H_{y,k_y} =  \sum_{\lambda=1}^{\Lambda-1} t_{y\lambda} e^{i\phi_\lambda(x)}\Psi^\dagger_{k_y,\lambda+1,s}  \Psi_{k_y, \lambda,s}\nonumber\\
&\hspace{40pt}+t_{y\Lambda} e^{i(\phi_\Lambda(x)- k_y a_y)}\Psi^\dagger_{ k_y, 1,s}  \Psi_{ k_y, \Lambda,s} + H.c.
\end{align}
As a result, the eigenfunctions of ${H}={H}_{x}+{H}_{y} \equiv \sum_{k_y} H_{k_y}$  factorize as
$e^{i k_y y}\psi_{k_y}(x)$. Further, we focus on the $x${\,} -dependence of  
$\psi_{k_y}(x)$ and treat $k_y$ as  a parameter.

\section{Resonant scattering and edge states \label{sec:hamil}}

In this section we consider a typical strip introduced above. As an example, we choose a strip with a supercell that is composed of four stripes, i.e. $\Lambda=4$, see Fig. \ref{fig:tb_model}. Moreover, we note that the effective Hamiltonian $H_{y,k_y}$ is diagonal in spin space, so we focus below only on one spin component  and suppress the spin index $s$ in this section.

To construct the effective model, we linearize the spectrum around the Fermi points, which allows us to treat the system analytically \cite{Two_field_Klinovaja_Stano_Loss_2012,MF_wavefunction_klinovaja_2012}.
From now on, we  focus on momenta close to the Fermi points $\pm k_F$ determined by the chemical potential $\mu$ which defines the Fermi level.
The annihilation operator $\Psi(x)$ acting on states being close to the Fermi points can be represented in terms of slowly varying right and left movers, $R_{\lambda} (x)$ and $L_{\lambda} (x)$, respectively,
$\Psi(x) =\sum_\lambda [R_{\lambda} (x)  e^{ik_F x} + L_{\lambda}  (x) e^{-ik_F x}]$.  The kinetic term $H_{x}$, the linearized approximation of Eq. (\ref{h_x}), is rewritten as
\begin{align}
&H_{x} =\int dx \sum_{\lambda=1}^4  \hbar \upsilon_F (-i R_{\lambda}^\dagger \partial_x R_{\lambda} +  i L_{\lambda}^\dagger \partial_x L_{\lambda} ),
\end{align}
where $\upsilon_F =\partial \epsilon/\partial (\hbar k_x)|_{k_F}$ is the Fermi velocity. The hoppings between neighbouring stripes [see Eq. (\ref{1d_y})] are relevant
only if a {\it resonant scattering condition} is satisfied: the magnetic field phase $\phi_\lambda (x)$ should be commensurable with $2 k_F x$. 
This condition enables a compensation of the phase difference $2 k_F x$ between left and right movers  by the magnetic phase $\phi_\lambda (x)$ during the hopping process. As a result, 
the right and left movers  get coupled and a gap in the spectrum can be opened that lowers the energy of the system.
From now on, until stated otherwise,  we consider a strip with the supercell consisting of uniformly distributed stripes, $a_{y\lambda}=a_y/4$, see Fig. \ref{fig:tb_model}b. In this case, all phases $\phi_\lambda (x)$ are equal to each other. Thus, the resonant scattering condition $\phi_\lambda (x)= 2k_F x$ is satisfied for the magnetic field $B_1 = 2 k_F \hbar c/e  a_{y\lambda}$.
The hopping part of the Hamiltonian $H_y$  is rewritten in the basis ${\bf \Psi} = (R_{1}, L_{1}, R_{2}, L_{2}, R_{3}, L_{3}, R_{4}, L_{4})$ as
\begin{align}
&H_y =\int dx 
 \Big[\sum_{\lambda=1}^3 t_{y\lambda} R_{\lambda+1}^\dagger L_\lambda  + t_{y4} e^{-i k_y a_y }R_{1}^\dagger L_{4}\Big]
+ H.c.,
\end{align}
where we have neglected all fast-oscillating terms.
The Hamiltonian density $\mathcal{H}$, defined via $H=H_x+H_y = \int dx\ {\bf \Psi}^\dagger(x) \mathcal{H} {\bf \Psi}(x)$, is given by
\begin{widetext}
\begin{equation}
\mathcal{H} =\begin{pmatrix}
\hbar \upsilon_F \hat k &0&0&0&0&0&0&t_{y4}e^{-i k_y a_y}\\
0&-\hbar \upsilon_F \hat k&t_{y1}&0&0&0&0&0\\
0&t_{y1}&\hbar \upsilon_F \hat k&0&0&0&0&0\\
0&0&0&-\hbar \upsilon_F \hat k&t_{y2}&0&0&0\\
0&0&0&t_{y2}&\hbar \upsilon_F \hat k&0&0&0\\
0&0&0&0&0&-\hbar \upsilon_F \hat k&t_{y3}&0\\
0&0&0&0&0&t_{y3}&\hbar \upsilon_F \hat k&0\\
t_{y4}e^{i k_y a_y}&0&0&0&0&0&0&-\hbar \upsilon_F \hat k  
\end{pmatrix},
\label{den_big}
\end{equation}
\end{widetext}
where $\hbar {\hat k} = -i\hbar \partial_x$ is the momentum operator with eigenvalue $\hbar k$.

The energy spectrum $\epsilon_{\lambda,\pm} = \pm\sqrt{(\hbar \upsilon_F  k)^2 + t_{y\lambda}^2}$ [referred further to as the bulk spectrum]  is independent of $k_y$.  Here, the positive (negative) sign corresponds to the part of the spectrum above (below) the gap.
The system is fully gapped, and the size of the gap $2\Delta_g$ is determined by the smallest hopping matrix element, $\Delta_g={\rm min}_\lambda \{t_{y\lambda}\}$.  We note that such an opening of gaps at the Fermi level due to resonant scattering lowers the system energy and is similar to a Peierls phase transition \cite{Peierls,Braunecker_Loss_Klin_2009}.

There are no states
inside the bulk gap $\Delta_g$ in an infinite two-dimensional system. In contrast to that, 
a strip of finite width $W$ (see Fig.~\ref{fig:tb_model})
can potentially host states localized at the edges, whose energies lie inside the bulk gap. To explore the possibility of such edge states, we consider a semi-infinite strip ($x\geq0$) and follow the method developed in Refs. \cite{Two_field_Klinovaja_Stano_Loss_2012,MF_wavefunction_klinovaja_2012}.
At $x=0$ we  impose vanishing boundary conditions  on the wavefunction, ${\boldsymbol \psi}_{k_y}(x)|_{x=0}\equiv(\psi_{k_y,1}, \psi_{k_y,2},\psi_{k_y,3},\psi_{k_y,4})|_{x=0}=0$, where $\psi_{k_y,\lambda} (x)$ 
is the wavefunction defined in the stripe $\lambda$, and   $\Psi_{k_y,\lambda}$ is its corresponding particle operator.

\begin{figure}[!t]
 \includegraphics[width=0.9\columnwidth]{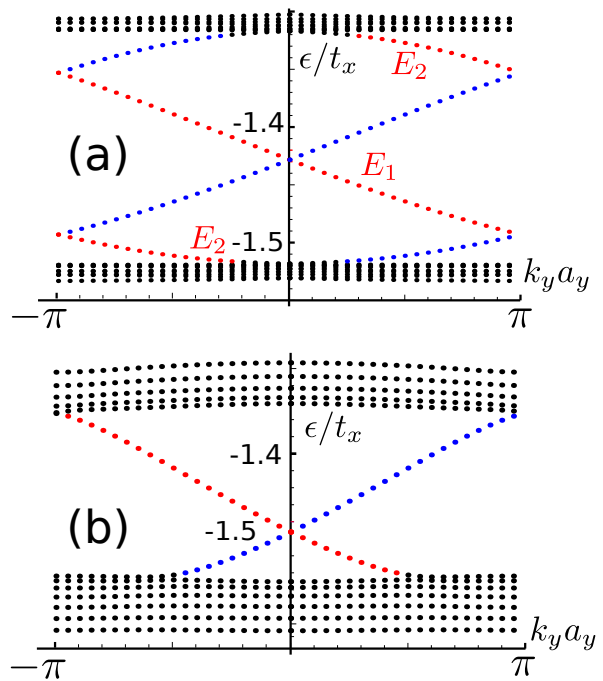}\\
 \caption{The spectrum $E(k_y)$ of the left  (red dots)  and of the right  (blue dots) edge states propagating along $y$ at the edges of a strip of width $W=203 a_x$ in the tight-binding model. The strip is at a quarter-filling, $\mu=-\sqrt{2}t_x$. The magnetic field $B_1$ corresponds to the resonant scattering condition, $\phi_\lambda (x)= 2k_F x$.  (a) The spectrum in the uniform case, $t_{y\lambda}\equiv t_y$, ($t_{y}/t_x=0.1$) is in  good agreement with analytical predictions $E_{1,2}(k_y)$ [Eqs. (\ref{dispersion_1})-(\ref{dispersion_2})]. (b) In the non-uniform case ($t_{y1}/t_x=t_{y2}/t_x=t_{y3}/t_x=0.3$ and $t_{y4}/t_x=0.1$), the edge states have their support in momentum space only in a reduced part of the first Brillouin zone. The edge states are topologically stable in the sense that they are robust  against all perturbations, as long as the gap remains open.
}
 \label{fig:spectrum}
\end{figure}

\begin{figure}[!b]
 \includegraphics[width=\columnwidth]{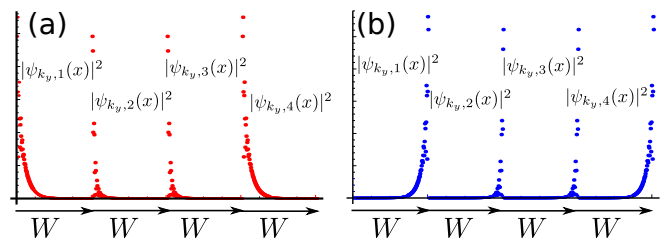}\\
 \caption{Plots of the probability density $|\psi_{k_y,\lambda}(x)|^2 $  as function of $x$ of (a) the left and of  (b) the right localized  states that decay exponentially away from the edge, obtained in the tight-binding model.  The parameters are chosen to be the same as in Fig. \ref{fig:spectrum}b, and $k_ya_y=\pi/8$. We note that, in contrast to the uniform case [see Eqs. (\ref{fun_1})-(\ref{fun_2})], the density patterns are different for different stripes. The exponential decay of the numerical simulations agrees well with that of the analytical solutions, Eqs.~(\ref{fun_1}) and (\ref{fun_2}).}
 \label{fig:states}
\end{figure}

In the uniform case of equal hopping matrix elements,  $t_{y\lambda}\equiv t_y$,  the spectrum of the left localized edge state is found to be given by
\begin{align}
&E_1(k_y) = -t_y \sin \left(\frac{k_ya_y}{4}\right), \label{dispersion_1} \\
&E_2(k_y)  = t_y {\rm sgn} (k_y) \cos \left(\frac{k_ya_y}{4}\right),\label{dispersion_2}
\end{align}
where the product $k_ya_y$ is defined inside the first Brillouin zone, $k_ya_y\in (-\pi, \pi]$. Similarly, the spectrum of the right localized edge state decaying at $x\leq0$ can be derived. We find that it is also given  by Eqs. (\ref{dispersion_1})-(\ref{dispersion_2}), however, with reversed sign, $E_{1,2}(k_y)\to E_{1,2}(-k_y)$ (see Fig. \ref{fig:spectrum}a). The wavefunction for the $E_1$ branch  is given by
\begin{equation}
{\boldsymbol \psi}_{k_y}^{(1)}(x) = \begin{pmatrix}
 1 \\   -i e^{i k_y a_y/4} \\ - e^{i k_y a_y/2} \\ i e^{3 i k_y a_y/4}
 \end{pmatrix}\sin (k_Fx) e^{-x/\xi_1},\label{fun_1}
\end{equation}
while for the  $E_2$ branch  by
\begin{equation}
{\boldsymbol \psi}_{k_y}^{(2)}(x) = \begin{pmatrix}
 1 \\    - {\rm sgn} (k_y) e^{i k_y a_y/4} \\  e^{i k_y a_y/2} \\ - {\rm sgn} (k_y) e^{3 i k_y a_y/4}
 \end{pmatrix}\sin (k_Fx) e^{-x/\xi_2},\label{fun_2}
\end{equation}
where the localization lengths  are given by $\xi_1 = \hbar\upsilon_F/\sqrt{t_y^2-E_1^2}$ and $\xi_2 = \hbar \upsilon_F/\sqrt{t_y^2-E_2^2}$. We note that
all left (right) localized edge states have negative (positive) velocities. This means that these edge states are chiral and the propagation along a given edge is possible only in one direction determined by the direction of the magnetic field $\bf B$. We note that in the uniform case $t_{y\lambda}\equiv t_y$ the energy spectrum can be mapped back to the extended Brillouin zone that is defined by a new lattice period $\tilde a_y=a_{y\sigma}=a_y/4$.  The energy spectrum of the left localized edge state $E_L(k_y)$ is given by
\begin{align}
E_L(k_y)= - t_y \cos (k_y \tilde a_y), \ \ k_y \tilde a_y \in (-\pi,0)
\label{SP1}
\end{align}
and of the right localized edge state $E_R(k_y)$ by
\begin{align}
E_R(k_y)= - t_y \cos (k_y \tilde a_y), \ \ k_y \tilde a_y \in (0,\pi).
\label{SP2}
\end{align}

In the non-uniform case of four distinct hopping matrix elements $t_{y\lambda}$  the calculation is more involved. For example, the energy spectum of the left localized edge state $\left|E(k_y)\right|<t_{y\lambda}$ is given implicitly by the following equation,
\begin{align}
e^{i(\vartheta_1+\vartheta_2+\vartheta_3+\vartheta_4)} = e^{-i k_y a_y},
\end{align}
where $e^{i\vartheta_\lambda}=\left(\sqrt{t_{y\lambda}^2-E^2}+ i E\right)/t_{y\lambda}$.
However, the  main features of the spectrum are similar to the one described in Ref. \cite{QHE_Klinovaja_Loss} for a strip with a supercell composed of two stripes. The main difference between the uniform and non-uniform case lies in the support of the edge states in  momentum space. In the non-uniform case, not for every value of $k_y$ there is a corresponding edge state, see Fig.~\ref{fig:spectrum}b. In addition, the probability density 
$|\psi_{k_y,\lambda}(x)|^2 $ is distributed non-uniformly among the four stripes of a supercell, see Fig. \ref{fig:states}.

We confirm our analytical results with numerical diagonalization of the tight-binding model, see Figs.~\ref{fig:spectrum} and~\ref{fig:states}. The Hamiltonians $H_x$ and $H_y$ are rewritten as
\begin{align}
&{\bar H}_{x} = \sum_{m, k_y, \lambda} (-t_x c^\dagger_{m+1, k_y,\lambda}  c_{m,k_y, \lambda} + H.c.),\\
&{\bar H}_{y} = \sum_{m, k_y} [\sum_{\lambda=1}^3 t_{y\lambda} e^{i\phi_\lambda(ma_x)}c^\dagger_{m, k_y,\lambda+1}  c_{m,k_y, \lambda}\nonumber\\
&\hspace{40pt}+t_{y4} e^{i(\phi_4(ma_x)- k_y a_y)}c^\dagger_{m, k_y, 1}  c_{m, k_y, 4} + H.c.],
\end{align}
where $a_x$ is the lattice constant in the $x$ direction. Here, we discretized the $x$ coordinate as $x=ma_x$ with $m$ being an integer.  The hopping matrix  element in the $x$ direction $t_x$ is chosen in such a way that the Fermi velocity at the Fermi level corresponds to $\upsilon_F$ defined above in the continuum model, $\upsilon_F =\partial \bar\epsilon_\lambda/\partial (\hbar k_x)|_{k_F}$, where $\bar \epsilon_\lambda = - 2 t_x \cos (k_x a_x)$ is the spectrum of ${\bar H}_{x}$ in $k_x$-momentum space. 

\section{Integer quantum Hall effect \label{IQHE}}

\subsection{IQHE for $\nu=2n$: spin-unpolarized system}

\begin{figure}[!tb]
 \includegraphics[width=\columnwidth]{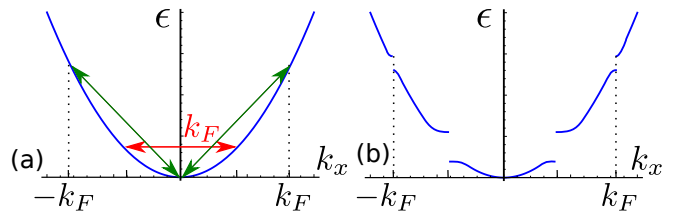}\\
 \caption{Sketch of the energy spectrum for $n=2$. (a) The resonant scattering condition $\phi_\lambda (x)= k_F x$ results in two possible scattering processes: direct (red line) and indirect (green line). (b) As a result, there are two gaps: one obtained in first-order perturbation expansion (in the hopping amplitude $t_{y\lambda}$) at $k_F/2$ and one obtained in second-order  at $k_F$. }
 \label{fig:nu_2}
\end{figure}

In the previous section we have considered a spinless strip. 
The system is gapped in the bulk but supports edge modes 
propagating along the edge in $y$ direction for the specific value of a magnetic field $B_1 = 2 k_F \hbar c/e  a_{y\lambda}$. The magnetic phases 
arising from this field are resonant, {\it i.e.}  $\phi_\lambda(x)=2k_Fx$,
allowing for resonant scattering between right and left movers. However, we should note that  magnetic fields $B_n=B_1/n$, corresponding to magnetic phases $\phi_{\lambda,n} (x)= 2k_F x/n$, also allow for resonant scattering, albeit in higher orders of  perturbation theory, see Fig. \ref{fig:nu_2}. Here, $n$ is a positive integer.

\begin{figure}[!tb]
 \includegraphics[width=\columnwidth]{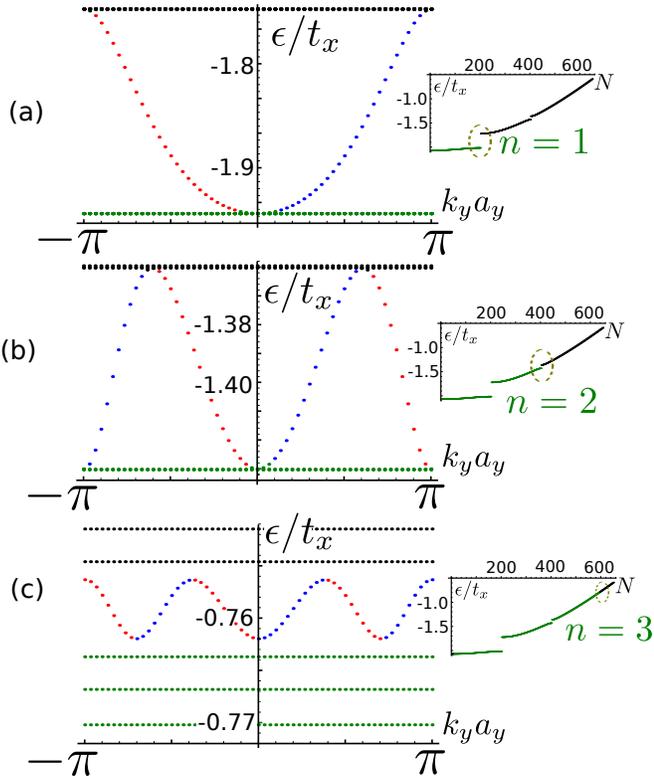}\\
 \caption{The spectrum $E(k_y)$ of left edge states (red dots)  and of right edge states (red dots) obtained in the tight-binding model. Inset: the bulk energy $\epsilon(N)$  corresponds to the $N$th energy level. The Fermi level separates filled (green dots) from empty (black dots) bulk states.  The supercell of the strip of width $W=1607 a_x$ consists of one stripe. 
The ratio between hopping matrix elements is chosen to be $t_y/t_x=0.4$.
 (a) In the case of  {\it direct} resonant scattering, {\it i.e.} $\phi_\lambda (x)=2 k_F x$ ($n=1$), there is one left and one right localized edge state.
 (b) [(c)] In the case of  {\it indirect} resonant scattering, {\it i.e.} $\phi_\lambda (x)= k_F x$ with $n=2$ ($\phi_\lambda (x)= 2k_F x/3$ with $n=3$), there are two (three) left and two (three) right localized edge states. All left (right) edge states propagate with negative (positive) velocities  along the strip. 
At the Fermi level and for a given edge,  all $n$ edge states  have about the same support and localization length, see Fig.~\ref{fig:functions}a for $n=2$. }
 \label{fig:edge_states}
\end{figure}

\begin{figure}[!t]
 \includegraphics[width=\columnwidth]{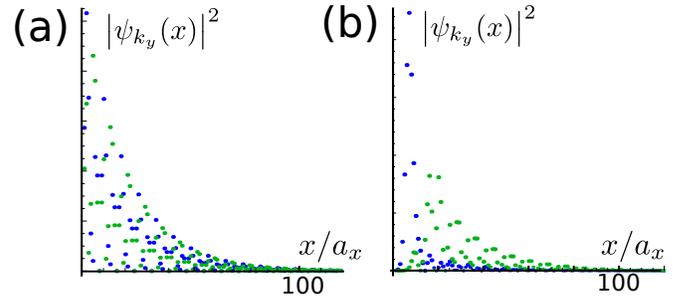}\\
 \caption{The probability density $\left| \psi_{k_y}(x)\right|^2$ of two left localized edge states obtained  numerically in the tight-binding model.  (a) The parameters are the same as in Fig. \ref{fig:edge_states}, corresponding to the hard-wall boundary conditions. Inside the second gap ($n=2$) at energy $\epsilon=-1.388 t_x$, there are two left localized edge states with different transverse momenta: one corresponds to $k_ya_y=0.8\pi$ (blue dots) and another one to $k_ya_y=-0.37\pi$ (green dots). We note that these two edge states are spatially not separated and have the same localization length. (b) In case of soft potential boundary conditions, $V(x)=0.001 t_x (x-30a_x)^2/a_x^2$ for $x<30a_x$, the localization lengths of two states at energy $\epsilon=-1.40 t_x$ with $k_ya_y=0.6\pi$ (blue dots) and $k_ya_y=-0.51\pi$ (green dots) are different.}
 \label{fig:functions}
\end{figure}

The effective (linearized) Hamiltonian can be obtained using a standard Schrieffer-Wolff  procedure \cite{BDL}.
For example, in case of a uniform strip with a supercell consisting of one stripe only, the Hamiltonian density is given by
\begin{equation}
\mathcal H^{(n)} = \begin{pmatrix}
                \hbar \upsilon \hat k & t_y^{(n)} e^{-i n k_y a_y}\\
                t_y^{(n)} e^{i n k_y a_y} &- \hbar \upsilon \hat k
               \end{pmatrix}
\end{equation}
 in the  lowest non-vanishing  order ({\it i.e.} $n$th order) in the hopping parameter $t_y$ [compare with Eq. (\ref{den_big})]. The size of the  gap $2\Delta_g =2 t_y^{(n)} \approx 2 t_y(t_{y\lambda}/\epsilon_F)^{n-1}$
is reduced  by the factor $(t_{y}/\epsilon_F)^{n-1}$, where $\epsilon_F$ is the 
characteristic Fermi energy in the system with $\epsilon_F \gg t_{y}$.  In addition, we note that gaps are also opened at momenta $\pm l k_F/n$, where $l$ is a positive integer that does not exceed $n$.  As a result, in total $n$ gaps are opened in the spectrum, and each of these gaps hosts right and left chiral edge states, see Fig.~\ref{fig:edge_states}. The spectrum of the edge states  $E^{(n)}_{ R/L}(k_y)$ inside the gap (opened by indirect $n$th order scattering processes) can be obtained from the  direct scattering process spectrum, $E_{R/L}(k_y)$ [see Eqs. (\ref{SP1})-(\ref{SP2})], just by rescaling the wavevector $k_y$,
\begin{align}
 E^{(n)}_{R/L} (k_y)=  E_{R/L} (n k_y).
\end{align}
As a consequence of such a scaling, the number of right-left edge state pairs is also increased by the same factor $n$. In other words, there are $n$ left and $n$ right localized edge states at the Fermi level. Our lowest-order analytical results are in good agreement with the exact numerical diagonalizations of the tight-binding model, see Fig.~\ref{fig:edge_states}.
We note that
all edge states  at a given edge have the same support and localization length in this approximation at the Fermi level and are not spatially separated from each other, see Fig.~\ref{fig:functions}a. 
This can be traced back to the hardwall condition imposed on our solutions. For soft-wall confinement the edge states get spatially separated, as we have confirmed numerically for the special case $n=2$, see Fig.~\ref{fig:functions}b.

So far, we have neglected the spin degeneracy present in the system and have focused only on one of the two spin components. Now, we  recall that the stripes are populated by both spin up and spin down particles. Because the spectrum is independent of spin (note that we neglect the Zeeman energy), it is straightforward to include it: Each state in the system can be occupied both by spin up and spin down, thus, for example, at each edge there are now in total $2n$ chiral edge states
at the Fermi level. As a result, the entire strip is spin-unpolarized.

The two-dimensional system in a perpendicular magnetic field $B$ is often characterized by the filling factor $\nu = N/N_L$, where
$N$ is  the total number of electrons in the system and $N_L$ the number of degenerate states in each Landau level.
Here, $N$ is defined via the particle density $\bar n = k_F/(\pi a_{y\lambda})$ as $N = 2 \bar n S$, where  $S$ is the area of the strip, and 
the coefficient of two accounts for the two spin directions \cite{Guiliani_Vignale}.
The degeneracy of the Landau level is given by  $N_L=eBS/hc$, where we do not take into account the spin degeneracy.
At a magnetic field $B_n$, the system is spin-unpolarized  and is at the filling factor $\nu=2n$.

\subsection{IQHE for $\nu=2n+1$: spin polarized system \label{seq_2n_1}}

So far, we  assumed  that states are spin degenerate, which restricted our consideration to even filling factors $\nu=2n$ only and it was not possible to obtain gaps at odd filling factors $\nu=2n+1$.
For example, the simplest odd filling, $\nu=1$, corresponds to a magnetic field $\tilde B_1=2 B_1$ and, thus, to the  phase $\phi_\lambda(x)=4k_Fx$.
The momentum difference arising in the hopping process, $\Delta k_x = 4k_F$, is too large to be absorbed by the right- and left-movers, so no gap opens. 

The question then arises whether the system can still lower its energy by opening a gap for odd filling factors. The answer is positive, and the mechanism works as follows.
If the stripes are rearranged twice as dense, $a_{y\lambda} \to \tilde a_{y\lambda}= a_{y\lambda}/2$, the  phase $\tilde \phi_\lambda(x) =\phi_\lambda(x) /2 =2 k_Fx$ 
can again lead to a resonant scattering between left  and right movers, assuming that the Fermi wavevector remains unchanged, {\it i.e.} $\tilde k_F= k_F$. This requirement must be satisfied together with the
requirement that
the number of stripes gets doubled at a fixed number of particles in the system.
It is  possible to satisfy these requirements only if the system is spin polarized, so that the spectrum is no longer spin-degenerate. As a result, there is effectively only one spin label, which is the same for all stripes. The last requirement finds its origin in the realistic assumption, adopted here, that the hopping between stripes preserves spin.

From now on we refer to such a spin-ordered state as to spin polarized. However, the developed spin order within a stripe can not only be the fully spin polarized one with all spins pointing uniformly into one  direction but also, more generally, a spin density wave (SDW)  with the spins rotating around some axis in spin space.
What kind of order is developed depends on the strength of electron-electron interactions (see discussion below in Sec. \ref{sec:order}).

\begin{figure}[!t]
 \includegraphics[width=\columnwidth]{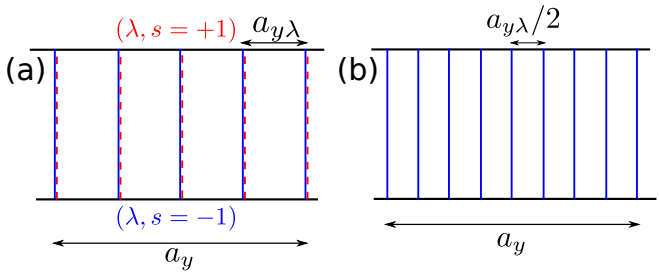}\\
 \caption{The supercell. (a) At the filling factors $\nu=2n$, the supercell consisting of $\Lambda=4$ stripes is spin-degenerate: each stripe is populated by spin up (red lines) and spin down (blue lines) electrons. (b) At the filling factor $\nu=2n +1$, it is energetically more favourable
for the same system to develop a supercell that consists of $2\Lambda=8$ spin polarized stripes, since this opens up resonant scattering that leads to energy-reducing gaps.}
 \label{fig:spin_polarized}
\end{figure}

As a consequence, instead of $\Lambda$ spin-degenerate stripes ($s=\pm1$) in the supercell, we work with  $2\Lambda$ stripes populated by spin polarized particles, see Fig. \ref{fig:spin_polarized}. Such a reconstruction of the system is favoured by energetic arguments. Indeed,  the kinetic energy is unchanged but an opening of the gap due to the resonant scattering lowers the total energy of the system. Hence, at the filling factors $\nu=2n+1$ the system gets spin polarized. The direction of the spin polarization can be determined, for example, from the requirement to minimize the Zeeman energy.

\section{Fractional Quantum Hall Effect \label{sec:order}}

\subsection{FQHE for $\nu=2n/m$: spin-unpolarized system }

We have addressed all IQHE filling factors in Sec. \ref{IQHE}. However, the scheme developed so far does not allow us to account for the FQHE filling factors. For instance, at the filling factor $\nu=2/m$ ($m>2$ is an integer), the magnetic phase in a spin polarized system, $\phi_\lambda (x)=  m k_F x$, is {\it larger} than the phase difference between left and right movers, which is given by $2k_Fx$.
Thus, resonant scattering is not possible because the momentum transfer in a hopping process, $2mk_F$, is too large to be absorbed by free  particles at the Fermi level. The mechanism that can accommodate such large momentum transfer is  {\it Umklapp scattering} \cite{Ashcroft_Mermin}. However, Umklapp processes emerge only in the presence of a periodic potential in real space. It is at this point where we invoke interaction effects: In this section we show that, within a stripe, interacting electrons  provide effectively such  periodic potentials for themselves,
with a period dependent on the Fermi wavevector $k_F$. This is again a mean field picture: A given electron at the Fermi level moves in the periodic potential created by all the other electrons of the stripe.
Our first goal  is to find the  period that is energetically most favourable in a given stripe.  In what follows we assume that this period is given by $8k_F$. Based on this, we will then be able to describe all filling factors for the FQHE, again with the help of resonant scattering but now assisted by Umklapp processes. In  Appendix \ref{Appendix} we provide semi-quantitative arguments based on interactions why $8k_F$ is the  dominant periodicity. These arguments are based on a mean field theory approach that takes into account explicitly electron-electron interactions in the Luttinger liquid framework. 
Our main goal here is to show that the stripe model shows gaps at the known fractional filling factors due to Umklapp scattering from $8k_F$ potentials. 
It remains an open question of deriving explicit values for the size of these gaps as a function of the systems parameters including interactions.

\subsubsection{Resonant filling factors $\nu=2n/m$}

We begin the discussion of the FQHE with the spin-unpolarized case, for which the two spin directions are independent, so that every energy level is two-fold degenerate in spin. The FQHE results from the Umklapp scattering process, occurring in the presence of the periodic potential $V (x) = V_0 \cos(Kx)$ with the wavevector $K=8k_F$,
 that involves higher Brillouin zones. (We recall here that, in contrast, the IQHE  takes place only inside the first Brillouin zone.) 
 Such an Umklapp mechanism giving rise to gaps was already discussed in Ref. \cite{QHE_Klinovaja_Loss}, so we only briefly comment on it here.

\begin{figure}[!t]
 \includegraphics[width=\columnwidth]{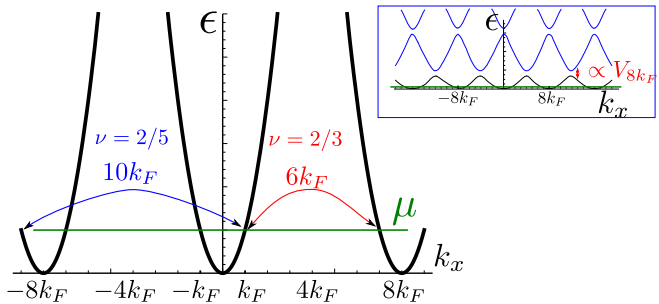}\\
 \caption{The spectrum of a stripe consists of spin-degenerate parabolas shifted by the reciprocal lattice vector $K=8k_{F}$, where the Fermi wavevector $k_F$ is determined by the chemical potential $\mu$. The resonant scattering caused by hoppings between stripes and a magnetic field leads to the opening of gaps. For example, at the filling factor $\nu=2/3$ ($\nu=2/5$) the magnetic field accounts for a change $6k_F$ ($10k_F$) in the momentum. The inset shows the first and second subbands due to the periodic potential with reciprocal lattice vector $K=8k_F$.  The size of the gap between two subbands  is determined by  $V_K$.
 }
 \label{fig:parabolas}
\end{figure}

Using the Bloch theorem, we arrive at the spectrum with the Brillouin zone of size $K=8k_F$ in reciprocal space, see Fig. \ref{fig:parabolas}. 
The hopping terms $t_{y\lambda}$, accompanied by the phase factor $\phi_{\lambda}$ containing the $B$ field, is now treated as a perturbation. 
Here, we again consider the uniformly distributed stripes in the supercell, resulting in equal phases $\phi_\lambda(x)\equiv\phi(x)$. 
In lowest order, the magnetic field
leads to a gap at the Fermi level (via resonant scattering) only if it results in magnetic phases commensurable with $2k_F$,
\begin{equation}
\phi(x) \equiv \frac{e}{\hbar c} B a_{y\lambda} x = \pm 2k_F \frac{p}{n}x + q K x.
\end{equation}
Here, $q$ is an integer, $n$ is a positive integer, and $p$ is a positive odd integer coprime to $n$ such that $p < 2n$. 
Such values of magnetic fields, {\it i.e.},
\begin{equation}
B=  2k_F \frac{\hbar c}{e a_{y\lambda} } \frac{4qn\pm p }{n},
\label{mag_field_resonance}
\end{equation} 
correspond to the filling factors  $\nu = 2n/m$ with $m=4 q n\pm p $. 
We note that $m$ is always an odd integer.

\subsection{FQHE for $\nu=(2n+1)/m$: spin polarized system}

The explanation of the FQHE constructed above for spin-unpolarized systems cannot account for  filling factors of the form $\nu=(2n+1)/m$, where $n$ and $m$ are both positive integers, and $m$ is odd. The resulting magnetic phases $\phi_\lambda (x) = (2n+1)k_F x/m $ do not have resonant values in this case. Thus, similarly to Sec. \ref{seq_2n_1},  we allow for the restructuring of the supercell in such a way that the number of the stripes is doubled and, as a result, the system gets spin polarized. As a further consequence, the magnetic phases are halved, ${\bar \phi}_\lambda(x)=\phi_\lambda (x)/2$, whereas the Fermi vector $k_F$ remains unchanged. 

In a next step, we should determine the (reciprocal) period of the  potential arising from the interaction-induced SDW in the one-dimensional stripe. Again, it could be $2k_F$, $4k_F$, $6k_F$,  $8k_F$, and so on. However, as shown above, $2k_F$ and $6k_F$ are less relevant than $4k_F$ and $8k_F$, and, moreover,
$4k_F$ (and $2k_F$) leads to an effectively time-reversal symmetric Hamiltonian at the resonant filling factors, and, as a result, the band gap is not fully developed. This would then lead to a higher energy state in comparison to a flat band gap generated by $8k_F$. Allowing for Umklapp processes, we obtain the resonant scattering condition for the spin polarized system,
\begin{equation}
\phi(x) \equiv \frac{e}{\hbar c} B \frac{a_{y\lambda}}{2}x = \pm 2k_Fx \frac{p}{n} + q K x.
\end{equation}
Thus, we  see that if the period is indeed given by $K=8k_F$, above condition results in the values of filling factors $\nu = n/m$, where $m =4qn \pm p $ is an odd integer. 

By comparing filling factors for the spin polarized and spin-unpolarized systems, we see that the even numerator $\nu=2n/m$ can be obtained in both cases. However, we note that for the spin polarized case  the gap opens in higher order perturbation theory, so the  gap size is smaller. This means that, generally, the unpolarized state is energetically favoured over the polarized one. However, if the Zeeman energy starts to play a significant role, the spin polarized state can have lower energy. We emphasize that the odd numerator filling factors $\nu=(2n+1)/m$ always lead to a spin polarized state.

We conclude this subsection with a few remarks about filling factors 
\begin{equation}
\nu=\frac{n}{m}\equiv \frac{n}{\alpha p + 4qn}.  
\end{equation}
First, the special values  $q=0$ (no Umklapp), $p=1$, and $\alpha=\pm 1$
correspond to the IQHE. 
Second, the choice of $p=1$ and $\alpha=1$ ($p=2n-1$ and $\alpha=-1$) with $q\neq0$
results in an integer number of flux quanta, $\Phi_0=hc/e$, attached to a particle, {\it i.e.},  $4q\Phi_0$ [$2(2q-1)\Phi_0$].
In this case, the attached flux is always an even number of flux quanta. This observation is  in full agreement with the composite fermion theory \cite{book_Jain} constructed for the filling factors $\nu = n/(1+2kn)$, where $k$ is a positive integer. Thus, we conclude that our stripe model supporting fermions that scatter over higher Brillouin zones is equivalent to composite fermions with attached fluxes.

Finally, we note that the FQHE can   be mapped back to the IQHE by redefining the electron charge $e$ as $e^\star=e/m$. This redefinition allows us to keep all scattering events inside the first Brillouin zone. However, one should be careful applying this mapping.
The mapping  from $\nu=n/m$ to $\nu=n$ applies  only for $\alpha=1$. In case of $\alpha=-1$, the hopping effectively takes place in the presence of a magnetic field applied in the opposite direction, $B \to -B$. As a result, the propagation direction of the edge states is opposite to the one at the IQHE filling factors. This again corresponds to the composed fermion theory in which an applied magnetic field $B$ and an enhanced magentic field $\tilde B$ point in opposite directions.

\subsection{Hall conductance}

\subsubsection{Streda formula}

The presence of edge states and a gap in the bulk spectrum  can be tested in transport experiments. For example, the Hall conductance $\sigma_H$ is expected to exhibit plateaus (as function of the magnetic field $B$) on the classical dependence curve $\sigma_H \propto 1/B$, if the Fermi level lies in the bulk gap.  The Hall conductance can be calculated using the Streda formula \cite{Streda},
\begin{equation}
\sigma_H = e c \left( \frac{\partial {\bar n}}{\partial B}\right)_\mu,
\end{equation} 
where ${\bar n}$ is the bulk particle density, which is uniquely determined by the Fermi wavevector. From now on, we are interested in the Hall conductance in the case where the chemical potential
lies in the gap. Thus, we focus on the values of magnetic fields close to those determined by the resonant scattering condition. 
The Fermi wavevector, at which a gap develops, is connected to the magnetic field via the relation [see, for example, Eq. (\ref{mag_field_resonance})]
\begin{align}
&k_F =\nu \frac{e}{4\hbar c} B a_{y\lambda}.
\label{B_kF}
\end{align}
If $\mu$ lies in the gap opened by the Umklapp scattering, the change in the density, $d{\bar n}$, due to a change in the magnetic field, $dB$, is given by 
\begin{equation}
 d{\bar n}  = \frac{d k_F}{\pi a_{y\lambda}}
=\nu \frac{e}{h c}  d B.
\end{equation}
Hence, the Hall conductance (in SI units) assumes the  FQHE plateaus,  
\begin{equation}
\sigma_H = \nu\frac {e^2}{h}
\label{Hall_conductance}
\end{equation}
at the filling factor $\nu$ and is independent of any microscopic system parameters. The width of the plateaus
is non-universal and determined by the $\,$ gap size $2\Delta_g \propto 2t_{yl} (t_{yl}/\epsilon_F)^{(n-1)}$, which depends on the filling factor $\nu$ via the integer $n$. 
Moreover, we have checked numerically that the bulk gap $2\Delta_g$ never closes for any finite ratio between $t_x$ and $t_{yl}$. This strongly suggests that 
the Hall plateaus at $\nu$ remain present  also  in the isotropic limit $t_x \sim t_{yl}$ (which cannot be treated by our methods).

We  note that the density ${\bar n}$ depends also on the chemical potential $\mu$, i.e. ${\bar n}={\bar n}(k_F(B),\mu)$. But if the chemical potential lies inside the gap, the density depends only very weakly
(finite size effect) on it via the edge states whose filling changes by changing $\mu$, whereas the density of the filled continuum below the gap does not change.

From the scenario of FQHE and IQHE developed above, it  follows that the IQHE is more stable against disorder than the FQHE. This is intuitively obvious, since the latter requires Umklapp scattering through higher Brillouin zones, and this
is only possible if there is a periodic potential due to the interaction-induced CDW (or SDW). However, disorder can easily destroy such periodic structures; therefore, it must be sufficiently small. A rough estimate is that
the disorder potential should be smaller than the gaps induced and that the disorder-induced mean free path is much larger than the period $2\pi/K$.

\subsubsection{Contrast to the TKNN formalism} 
 
We emphasize that it is crucial for obtaining the FQHE plateaus in our model that the size of the Brillouin zone $K$ depends on $k_F$, so that $B \propto k_F$ in Eq.~(\ref{B_kF}). 
 This is in stark contrast to the case of an external periodic potential with a {\it fixed} period,  a problem treated by the famous Thouless-Kohmoto-Nightingale-den Nijs (TKNN) formalism \cite{TKNN}.
 In particular, in the TKNN case the periodic potential is independent of $k_F$, 
and thus $B \propto k_F$ (at resonance) does not hold in this approach.

Above, we calculated the Hall conductance via the Streda formula. Alternatively, one might also attempt to use the Kubo formalism  like in the TKNN approach. However, in this case, one needs to be aware
of the fact that, in a non-equilibrium transport situation, the Fermi wavevector $k_F$ depends in general on the chemical potentials of the incoming ($\mu_{in}$)  and outgoing ($\mu_{out}$) current leads attached to the strip, {\it i.e.} $k_F=k_F(\mu_{in}, \mu_{out})$. Similarly, if we consider an infinite strip in the presence of an electric field described by a time-dependent gauge potential ${\bf A}_0(t)$, the Fermi wavevector will depend on it,  {\it i.e.} $k_F=k_F({\bf A}_0(t))$.
 As a consequence, a periodic potential that depends on $k_F$ (like in our model) will  also depend on those external driving fields. In particular, such a dependence will give rise to an extra-contribution in linear response calculations in addition to the standard Kubo response function. In other words, for the stripe model considered here, the Hall conductance is in general no longer simply given by the Chern number of the corresponding lattice model described by TKNN.
 Thus, we conclude that our stripe model with an effective periodic potential, induced by interactions, is fundamentally different from  non-interacting two-dimensional lattice models with fixed lattice parameters \cite{Hofstadter,TKNN,Thouless}, which exhibit only the IQHE but not the FQHE \cite{TKNN,Thouless,Avron}.

\section{ Even denominator FQHE \label{sec:even}} 

\subsection{Non-uniform supercells}

\begin{figure}[!tb]
 \includegraphics[width=\columnwidth]{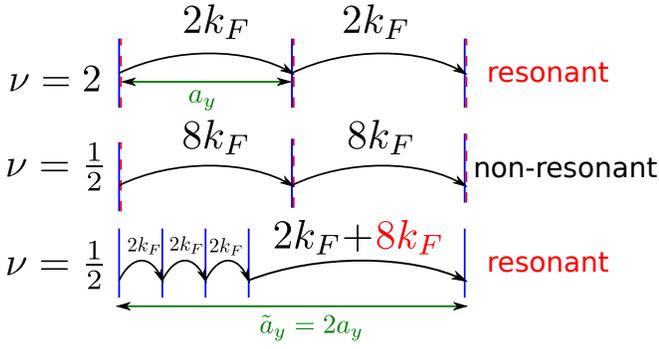}\\
 \caption{Resonant scattering at filling factors $\nu=2$ and $\nu=1/2$. The system is spin-unpolarized at $\nu=2$, so stripes are degenerate in spin: spin up (red lines) and spin down (blue lines). At $\nu=1/2$, the initial supercell of size $a_y$ would be non-resonant and no gaps would be opened. Thus, it is energetically more favourable for the system to restructure into a supercell with
a non-uniform distribution of  stripes  and with doubled size $\tilde a_y = 2 a_y$, so that resonant scattering again becomes possible. 
 Moreover, the system becomes spin polarized at $\nu=1/2$.}
 \label{fig:phases}
\end{figure}

The model for FQHE developed above predicts gaps and thus plateaus in the Hall conductance only at  filling factors with odd denominator. 
However, filling factors with even denominators, such as $\nu=1/2$ or $\nu=5/2$ have not been addressed so far. To deal with such values, we   allow now also for {\it non-uniform}
distributions of stripes inside a supercell.

Let us first focus  on the particular example of $\nu=1/2$. We consider a supercell of size $a_y$ initially composed of one stripe that can be occupied by both spin up and spin down electrons, see Fig. \ref{fig:phases}. At the filling factor $\nu=1/2$, the hopping in the $y$ direction results in the momentum transfer $8k_F$, which does not lead to  resonant scattering between right and left movers. To induce such scatterings, the stripe positions should be changed. First, the supercell gets doubled in size by merging two neighbouring supercells. Second, the system becomes spin polarized, so that there are in total  four stripes in the enlarged supercell. The separation between stripes is determined by the requirement to generate  resonant scattering. We immediately see that the new stripe positions are given by $a_{y1}=a_{y2}=a_{y3}=a_{y4}/5=a_y/4$, and
the corresponding magnetic phases become 
\begin{align}
&\phi_1 (x)=\phi_2 (x)=\phi_3 (x)=2k_Fx, \nonumber \\ 
&\phi_4 (x)=10k_Fx\equiv2k_Fx + Kx.
\end{align}
These phases, in turn, lead to the desired resonances that eventually open up gaps, see Sec. \ref{sec:hamil}. Here, the excess phase $Kx$ is absorbed by the periodic potential $\cos (Kx)$ via the Umklapp scattering.
The resonant scattering is again of direct type (first order in $t_{y\sigma}$), as shown in Fig.~\ref{fig:parabolas} for $\nu=2/3,2/5$.

Next, we generalize this scenario to other even denominator filling factors.
The size of a new supercell and new stripe arrangements can be found, again, from the resonant scattering condition. The basic idea is to merge several initial supercells and to rearrange stripes in such a way that {\it all} magnetic phases become resonant. In particular, for  filling factors of the form $\tilde\nu=l/2$, where $l$ is an odd integer, we develop a 
scenario where we map the effective Hamiltonian back to the  IQHE regime at filling factor $\nu=l$. First, we compose a new supercell of size $\tilde a_y = Ma_y$ from $M$ initial supercells consisting of one spin-unpolarized stripe. This results in $\tilde \Lambda=2M$ stripes for the spin polarized system. We note that if we reduce the distances between all stripes in the supercell by a factor of four, giving 
$\tilde a_{y\lambda} = a_{y\lambda}/4$ with $\lambda=1, ... ,\tilde\Lambda-1$, the magnetic phases $\phi_\lambda(x)$ become resonant. The only condition left is to make also $\phi_{\tilde\Lambda}(x)$ resonant. 
 Obviously, 
the choice $\phi_{\tilde\Lambda}(x) = \phi_1(x) + K x$ is resonant. As a result, we arrive at $M=2l$.
For example,  for $\nu=5/2$  we get $M=10$, {\it i.e.} the supercell  consists of $\tilde \Lambda=20$ spin polarized stripes.

We have just shown how to generate filling factors $\nu$ with denominator equal to $2$. Next, we generalize this  to arbitrary even denominators. The resonant magnetic  phases $\phi_{\lambda}$ are equal to $2k_Fx/\nu_{\lambda}$, where $\nu_\lambda=n_\lambda/m_\lambda$ is one of the odd denominator FQHE filling factors with $m_\lambda= \alpha_\lambda p_\lambda + 4q_\lambda n_\lambda$, where $\alpha_\lambda=\pm1$ (see above). The distances between stripes in the supercell are chosen such that each magnetic phase $\phi_\lambda $ is resonant,  and the sum of all magnetic phases over the supercell determines an even denominator filling factor $\tilde\nu$. This results in the relation $\sum_{\lambda=1}^{2M} \phi_\lambda = 4M{k_Fx}/{\tilde{\nu}}$, giving the following
Diophantine equation,
\begin{equation}
\sum_{\lambda=1}^{2M} \frac{1}{\nu_\lambda} = \frac{2M}{\tilde{\nu}}.
\label{dioph}
\end{equation}
This equation is necessary, but, unfortunately, not always sufficient. For example, in case of a supercell composed of $2M=2$ stripes, the filling factors $\nu_1=1$, $\nu_2=1/3$, and $\tilde{\nu}=1/2$ satisfy Eq. (\ref{dioph}) but do not lead to a fully gapped system: gaps open only for half of the branches. As a result, such a restructuring is  energetically less favourable than one with all branches being gapped. 

We further note that the Diophantine equation, Eq.~(\ref{dioph}), not always has a unique solution. For example, the filling factor $\tilde\nu=l/2$, where $l$ is an odd integer, can also be mapped to the effective Hamiltonian corresponding to the filling factor $\nu=[l/2]+1$, where $[x]$ denotes the integral part of $x$. 
For this, we choose $\nu_\lambda = \nu$ for $\lambda=1, ... ,2M-1$ and $\nu_{2M} = \nu/(1+4\nu)$.  In this case, the number of stripes is equal to $2M= 4 \nu \tilde\nu/(\nu-\tilde\nu)=4l([l/2]+1)$. In particular,  for $\nu=5/2$  we get $M=30$, {\it i.e.} the supercell  consists of $\tilde \Lambda=60$ spin polarized stripes in this case (instead of 20 obtained before). In other words, after every 60th stripe there is a shift in distance. Viewed this way, the new arrangement is still highly regular.

We note that if several solutions are possible, the one with the smallest value of $n_\lambda$ in $\nu_\lambda$ should be preferred, because this leads to the opening of gaps in lowest order, and, as a result, to  larger gaps. Moreover, the scenario with the smallest value of $M$ corresponds to the less drastic restructuring of the system.

In close analogy to the odd denominator FQHE discussed above, we can define an effective charge $e^\star$ for filling factors $\tilde\nu=l/2$. In both stripe scenarios considered above the mapping is done from $\tilde\nu$ to the IQHE filling factor:
to $\nu=l$ in the first scenario  and to $\nu=[l/2]+1$ in the second scenario. This corresponds to an effective charge $e^\star=e/2$ in the first scenario and to an effective charge $e^\star=e l/(l+1)$ in the second scenario.
Alternatively, we can define an effective charge $\tilde e^\star$ as an excess charge carried by edge channels in addition to the total charge carried by  all edge states of the corresponding $[l/2]$ fully filled Landau levels (where each of them carries a charge \cite{book_Jain} $e$). We note that for odd values of $l$ such an effective charge is given by $\tilde e^\star=e/2$ in both stripe scenarios.

Finally, we note that, independent of the stripe distribution, the Hall conductance is again given by Eq.~(\ref{Hall_conductance}), but now for even denominator filling factors $\nu$.

\subsection{Energetics of stripes:  uniform {\it vs.} non-uniform distributions}

The reconstruction of a supercell  opens gaps in the spectrum and lowers the total energy. However, such a redistribution of stripes also leads to an increase in  interaction energy between particles. 
Whether a reconstruction occurs or not depends on  which contribution to the energy is larger: the energy gain from  opening of  gaps or the energy cost of bringing interacting particles closer to each other. 
We will not be able to decide this question here. Still, it is useful to  formalize the problem a bit further as follows.
The opening of a gap $\Delta_g$ leads to the energy gain of the well-known Peierls-form \cite{Peierls},
\begin{align}
\Delta_E^{(g)}\approx \frac{S \Delta_g^2}{\hbar \upsilon_F a_{y}} {\rm ln} \frac{\Delta_g}{D},
\end{align}
where $D\gg\Delta_g$ is an energy cut-off. To estimate the energy associated with the reconstruction, we model the stripes as a system of charged
cylindrical wires of radius $\rho$, separated by a distance $a_y$. The energy of such a model is then given by \cite{Morse}
\begin{equation}
V_{pot} =S (e\bar n)^2 a_y {\rm ln} \frac{a_y}{2\pi\rho}.
\end{equation}
As a consequence, the ratio of gap energy $\Delta_E^{(g)}$ to potential energy $V_{pot}$, given by
\begin{equation}
 \eta \approx  \frac{ \Delta_g^2}{\hbar \upsilon_F a^2_{y} (e\bar n)^2} \frac{{\rm ln} (\Delta_g/D)}{\alpha_{pot}},
\end{equation}
determines whether the reconstruction is $\,$ energetically 
favourable ($\eta >1$) or not ($\eta<1$). 
Here, 
$\alpha_{pot}=(E_{pot,1}-E_{pot,2})/ (e\bar n)^2 a_y$  compares 
 the two potential energies associated with a uniform and  non-uniform distribution of stripes, respectively.
In general, to compare analytically all energy gains and losses  is quite a challenging task and beyond the scope of the present work.
At this point, it seems that the answer to the question of restructuring can only be found by ab-initio numerics. At the same time, we can argue that there are some optimal even denominator filling factors to be observed.
On the one hand, for large filling factors, so at weak magnetic fields or at high densities, the periodic structures are not yet well-developed due to the lack of strong interactions. On the other hand, for small filling factors, so at strong magnetic fields or at low densities,  the energy gain from the restructuring is small. This is connected to the fact that the size of the opened gap is proportional to the hopping matrix  element $t_{y\lambda}$ that is small due to larger distances between stripes. This can shed some light on the fact that only one of the even denominator filling factors $\nu=5/2$ is observed \cite{5_2_exp_1,5_2_exp_2}.

As predicted by our theory, the size of a supercell is  largest in case of even denominator filling factors, so it is easier to observe the stripe periodicity along a strip in this case.
We believe that it is worth the effort to test our hypothesis based on self-organization of particles into periodic structure in real systems. A most suitable candidate for such an experiment is graphene where the surface can be accessed directly by local probes \cite{Amir_graphene}. According to our theory, the underlying periodic structure should be uniform for odd denominator  FQHE, 
whereas it is  non-uniform for even denominator FQHE.
Our predictions can also be tested in optical lattices \cite{Lewenstein} that allow for high control of lattice parameters and of the equivalent of magnetic phases \cite{Demler_Lukin_2005,Demler_Lukin_2007,Daghofer_2012,Cooper_Dalibard_2013,Yao_Lukin_2013}.

\hspace{1pt}

\section {Conclusions \label{sec:conc}} 
In this work we have studied topological properties
of an infinite anisotropic strip in the quantum Hall regime.
The edge states in the induced  gap are identified for both the IQHE and FQHE filling
factors. These edge states can be considered topologically stable in the sense that
perturbations that are weak compared to the induced gap will not affect the edge states
in an essential manner, they still are localized at one edge and stay chiral. 
Moreover, we checked numerically that the gaps never close for any finite ratio between 
hopping in $x$ and $y$ direction, lending further support to the topological stability of the 
results derived in the anisotropic regime, in particular of the Hall conductance, $\sigma_H=\nu e^2/h$.

We discussed the connection between composite
fermion theory built on particles with attached flux
and Umklapp scattering processes involving higher Brillouin
zones. Moreover, we address the even denominator
filling factors and demonstrate that restructuring
of sites inside the supercell can lead to the opening
of gaps in the spectrum. However, whether such a restructuring
is energetically favourable or not depends on non-universal system
parameters. To address this issue, it would be desirable to perform
ab-initio numerics on the interacting electron gas and see if opening of gaps
combined with a Peierls transition (with dimerization of the CDW or SDW within a stripe)
will lead to a stripe formation of the type assumed in this work.
Similar numerical calculations have been performed before \cite{book_Jain,Wigner_Girvin,Wigner_Kivelson,Halperin_crystall,anisotropy_CDW,Wigner_Jain},
but without allowing for a Peierls transition that could give the necessary energy gain to favour stripe formations in the quantum Hall regime of a standard 2DEG.
Furthermore, it would be desirable to search experimentally for such non-uniform density distributions, similar to recent experiments at high Landau levels \cite{anisotropy_CDW,Wigner_Jain,anisotropy_West,QHE_strips,strips_Review}.
Finally, given the high control in optical lattices \cite{Lewenstein}, the stripe model introduced here might be directly
 implemented  with cold atoms and molecules, where the equivalent of magnetic field effects can be
efficiently simulated.
\cite{Demler_Lukin_2005,Demler_Lukin_2007,Dalibard_RMP_2011,Daghofer_2012,Cooper_Dalibard_2013,Yao_Lukin_2013}

As an outlook we would like to mention that the stripe model presented here opens up the
possibility to access theoretically a variety of other physical quantities such as charge and spin
susceptibilities in the presence of electron-electron interactions, and perturbatively modified by the
hopping between the stripes. It will be interesting to see if this can lead to further predictions
that allow experimental tests of the underlying stripe model directly or indirectly.
Finally, for a complete characterization of the stripe model it will be necessary to go beyond the ground state
properties considered in the present work. In particular, it will be interesting to study also the excitations of the
stripe model and to see how far it is possible to make
contact with properties well-known  from the traditional FQHE in 2DEGs such as fractionally charged quasiparticles, braiding statistics, and Laughlin states \cite{QHE_Review_Prange,book_Jain}.  In addition, we have focused on the disorder-free systems, which is a good assumption for the optical lattice and cold atoms systems, but definitely  not for 2DEGs. Thus, it would require an additional study of effects of disorder that, in principle, could interfere with creation of the periodic potential. These questions are beyond the scope of this work and will be addressed elsewhere.

{\it Acknowledgments}. This work is supported by the Swiss NSF and NCCR QSIT. We would like to thank G. M. Graf, B. I. Halperin, T. Meng, and P. Stano for helpful discussions.

\appendix

\section{Scattering potential $8k_F$ \label{Appendix}}

It is well-known that electron-electron interaction effects get strongly enhanced at high magnetic fields (due to a suppression of the kinetic energy by the magnetic field),  and 
electrons tend to order into periodic structures due to the interactions between the electrons themselves.~\cite{book_Jain,Wigner_Girvin,Wigner_Kivelson,Halperin_crystall,anisotropy_CDW,Wigner_Jain,anisotropy_West,QHE_strips,strips_Review} 
So far we have assumed such periodic structures  along the $y$ direction, giving rise to the formation of stripes and supercells. Next, we allow for periodic structures due to interactions also along the $x$ direction, {\it i.e.} within a one-dimensional stripe, described as interacting Luttinger liquid \cite{Giamarchi}.
It is well-known that in such systems interactions generate a charge-density wave (CDW) or a spin-density wave (SDW) at some dominant Fourier mode with wavevector $K$ for the electrons around the Fermi surface.
In a mean field approach we assume that such  periodic modulations provide a periodic potential seen by an electron at the Fermi surface,  formally described by $V(x)=V_K \cos (K x)$ with period $a_x(k_F)=2\pi/K$ and amplitude $V_K$, 
where the reciprocal lattice wavevector $K$  depends on $k_F$. The next goal is  to determine the dominant Fourier mode and show that it is given by $K=8k_F$.

To begin with, we note that the CDW is described by density-density correlations of the form,
$\left<\rho(x)\rho(0)\right>\propto\cos(2lk_Fx)$, with $l$ being a positive integer \cite{Giamarchi}. Thus, we can get periodic potentials of the form $\cos(2k_Fx)$, $\cos(4k_Fx)$, $\cos(6k_Fx)$, $\cos(8k_Fx)$, {\it etc.} Intuitively, the dominant Fourier contribution should come with the smallest wavevector $2k_F$. This is, indeed, true for non-interacting or weekly interacting systems. However, the situation is quite different for strongly interacting systems, where the presence of a backscattering term $g_{1\perp} \cos (\sqrt{8}\phi_s)$ plays a crucial role. Here, $g_{1\perp}$ is a backscattering parameter, and $\phi_s$ is a boson field describing spin excitations in the Luttinger liquid representation \cite{Giamarchi}.

To find out which Fourier component of a periodic potential dominates, we study  the scaling dimensions (in a renormalization group sense) of the corresponding terms. We find that the 
 components with $l$  being even   scale differently from the ones with $l$ being odd due to the
presence of the  backscattering term.
This term and its higher harmonics  $ \cos (r\sqrt{8}\phi_s)$ (generated in $r$-th order perturbation expansion in the interaction~\cite{schulz})
 lead to a  slower decay of the $l$-even contributions. More quantitatively,
the $l$-even contributions of $\left<\rho(x)\rho(0)\right>$
scale as
\begin{equation}
\cos (2 l k_F  x) \left(\frac{\alpha}{|x|}\right)^{K_\rho l^2  },
\end{equation}
whereas the $l$-odd contributions scale as 
\begin{equation}
\cos (2 l k_F x) \left(\frac{\alpha}{|x|}\right)^{K_s + K_\rho l^2}.
\end{equation}
Here, $K_\rho$ ($K_s$) are the Luttinger liquid parameters for the charge (spin) sector, and $\alpha$ is a short-distance cut-off. For spin isotropic repulsive interactions $K_s$ is close to one, $K_s\approx 1$ \cite{Giamarchi,schulz}.
In contrast,  $K_\rho$ can be much smaller than one in a system  with long-range electron-electron interactions, $K_\rho\ll1$.  As a result, the $2k_F l$ component with  $l$ even dominates over the $2k_F (l-1)$ component for $K_\rho<1/(2l-1)$. This allows us to exclude, in particular, the $2k_F$ and $6k_F$ components from further considerations and to focus on the $4k_F$ and $8k_F$ components. 

We note that in spite of the fact that we have focused above on the CDW and, correspondingly, have considered the scaling dimensions of the density-density correlations, the same analysis can be carried out for a spin-density wave (SDW) \cite{Giamarchi}.
As a result, we again arrive at the conclusion that for  $K_s=1$  the $4k_F$ and $8k_F$ Fourier components are  the dominant ones.
 
\begin{figure}[!b]
 \includegraphics[width=0.9\columnwidth]{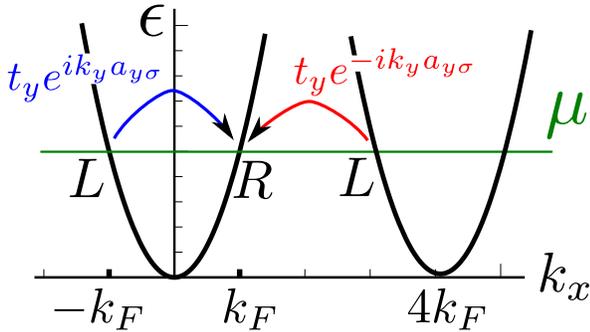}\\
 \caption{The reciprocal space of a stripe in the presence of an effective periodic potential with period $2\pi/4k_F$ consists of parabolas periodically shifted by the reciporcal lattice vector $4k_{F}$, where the Fermi wavevector $k_F$ is determined by the chemical potential $\mu$ (green line). The right ($R$) and left ($L$) movers are coupled in two ways by the hopping amplitude $t_y$: hopping in the positive direction (blue arrow) and in the negative direction (red arrow) along the $y$ axis. These hoppings carry opposite phase factors that lead to destructive interference, so that the system remains gapless and effectively time-reversal invariant in the presence of a magnetic field (not shown). }
 \label{fig:parabolas_4k_f}
\end{figure} 
 
Following the same scaling arguments, one might conclude that the $4k_F$ component [giving rise to a periodic potential $\propto \cos (4k_F x)$]
dominates over the $8k_F$ component [giving rise  to a periodic potential $\propto \cos (8k_F x)$].
However,  one important ingredient in this  argument is missing so far: the magnetic field. In the presence of the magnetic field, the resonant scattering between right and left movers becomes possible. This scattering leads to gaps in the bulk spectrum and thus lowers the  energy of the system similar to a Peierls phase transition. It is for this reason that a potential with $4k_F$ periodicity  becomes energetically less favorable than  a potential with $8k_F$ periodicity as we argue next.

As stated before, in the presence of a periodic potential, Umklapp processes, in which the periodic lattice can absorb the momentum $K$, becomes possible.  This leads to opening of gaps for a wider range of resonant magnetic fields (see below for details). However, we note that the periodic potential of the form $\cos(4k_Fx)$ is a very special one. Despite the fact that it is the magnetic field that induces resonant scattering via  higher Brillouin zones, the effective Hamiltonian defined around the Fermi level and containing $\cos(4k_Fx)$  stays effectively time-reversal invariant. 
This stems from the fact that one cannot distinguish  scatterings with  momentum $2k_F$ from the ones with a momentum  $-2k_F\equiv 2k_F+K_{4k_F}$, where $K_{4k_F}=4k_F$ is the reciprocal lattice wavevector. Thus, one cannot distinguish $\bf B$ from $-\bf B$. As a result, edge states, whose propagation direction is determined by the direction of the applied magnetic field $\bf B$, cannot exist in such an effectively time-reversal invariant system. We note in passing that, similarly, a potential with $2k_F$ periodicity results in an effectively time-reversal invariant Hamiltonian at resonant values of filling factors.

To demonstrate that the $8k_F$ scenario is energetically more favourable than the $4k_F$ one, we restrict ourselves from now on to the uniform case $t_{y\lambda}= t_y$ and to the filling factor $\nu=1$. This simplifies the analytical calculations (the supercell consists now only of one stripe) without loss of generality. The $8k_F$ scenario leads to  a uniform bulk gap that is independent of the transverse momentum $k_y$, $\Delta_{g, 8k_F}=t_y$ (see Fig. \ref{fig:spectrum}a). As mentioned above, the $4k_F$ scenario is effectively time-reversal invariant, and the right and left movers are coupled  via two hopping amplitudes that interfere, see Fig. \ref{fig:parabolas_4k_f}. This leads to the following change in the hopping part of the effective (linearized) Hamiltonian: $t_y e^{ik_y a_{y\lambda}}\to t_y e^{ik_y a_{y\lambda}}+t_y e^{-ik_y a_{y\lambda}}$ [compare with Eq. (\ref{den_big})]. The bulk gap becomes now $k_y$-dependent, 
\begin{equation}
\Delta_{g, 4k_F}= 2 t_y |\cos (k_y a_{y\lambda})|,
\label{g_4}
\end{equation}
and the system is gapless, specifically  at 
$k_y a_{y \lambda} =\pi/2$ and $k_y a_{y\sigma} =3\pi/2$. To 
compare energy gains for the $4k_F$ and $8k_F$ scenarios, respectively, we follow the standard calculation for the Peierls gap.~\cite{Ashcroft_Mermin} The free spectrum is given by 
$\epsilon(k_x) = \hbar^2 k_x^2/2 m_e + \mu$
and is independent of $k_y$. If resonant scattering at the wavevector $k_r=2k_F$ is induced, the spectrum becomes in leading order in $t_y$
\begin{align}
&\epsilon_{\pm,\kappa} (k_x, k_y) = \frac{\epsilon(k_x) +\epsilon(k_x +k_r)  }{2} \nonumber \\
&\hspace{25pt} \pm\sqrt{\left[\frac{\epsilon(k_x) -\epsilon(k_x +k_r)  }{2}\right]^2 + \Delta_{g,\kappa}^2(k_y)},
\end{align}
where $\epsilon_{-,\kappa}$ ($\epsilon_{+,\kappa}$) is the lower (upper) part of the spectrum and $\kappa=4k_F,8k_F$.
The total energies  $E_{4k_F}$ and $E_{8k_F}$, respectively, of the system are given by the integral over  all filled states in the first Brillouin zone.
The sign of the difference between these two energies is determined by the following expression,
\begin{align}
&E_{4k_F} - E_{8k_F} =\nonumber\\
&\hspace{30pt}\frac{S}{(2\pi)^2}\int_{-k_F}^{k_F} dk_x \int_{-\pi/a_{y\lambda}}^{\pi/a_{y\lambda}} dk_y\ [\epsilon_{-, 4k_F}-\epsilon_{-, 8k_F}]\nonumber\\
&=\frac{St_y}{\pi^2a_{y\lambda}}\int_{-k_F}^{k_F} dk_x\   \Big[ \sqrt{4+ {\delta \epsilon}^2} El\left(\frac{4}{4+\delta \epsilon^2}\right)\nonumber\\
& \hspace{120pt}- \frac{\pi}{2} \sqrt{1+ \delta \epsilon^2} \Big] >0,
\label{g_4-8}
\end{align}
where $\delta \epsilon =[\epsilon(k_x) -\epsilon(k_x +k_r)] /t_y$ and $El(x)$ is the elliptic integral. By plotting the function in the square bracket we see that the integrand stays always positive.
Thus, the  $4k_F$ scenario has always  higher energy than the $8k_F$ scenario.
In other words, the energy gain is smaller for the $4k_F$ case, where the system stays gapless, than for the $8k_F$ case, where the bulk gap is fully developed (and does not depend on $k_y$). 

 There are a few comments in order about the self-consistency of our mean field approach. 
The estimates of the Peierls energy performed above are 
valid in the limit of $t_y$ small compared to the  Fermi energy $\epsilon_F$ and thus also to the bandwidth
of the partially filled lowest subband. In this limit, the strength of the induced potential $V_K$ does not determine the Peierls energy gain [it is determined by $t_y$, see Eqs. (\ref{g_4}) - (\ref{g_4-8})] but it does determine 
the gap between the lowest and the next subband, see the inset of Fig. 9. 

However, the electron-electron interactions tend to renormalize $t_y$ and to increase it. 
As a result of this renormalization, the Peierls gap grows, leading to more energy gain, until the initial assumption of its smallness breaks down (and is no longer self-consistent). 
This breakdown happens when the Peierls gap
reaches the  size of the lowest subband (determined dominantly by the period $K$ and subdominantly by the strength of the potential $V_K$). 
Importantly, the bandwidth (of the lowest band) induced by $V_{4k_F}$  is (four times) smaller than the one induced by $V_{8k_F}$ (for a parabolic dispersion). 
Thus,  the maximum Peierls gap for the $V_{4k_F}$ case is presumably also (four times) smaller than the maximum one  for the $V_{8k_F}$ case. 
Importantly, the larger $V_K$  the sooner the associated RG flow of $t_y$ stops. In other words, from this consideration we obtain a self-consistency 
condition that would determine the value of $V_K$ as the one where the energy gain from the Peierls gap is maximal. 

We can see that there is a trend to a ``saddle point approximation'' physics: a very weak potential $V_K$ tends to become stronger to ensure 
the opening of the Peierls gap (by enabling Umklapp scattering),  but a very strong potential $V_K$ tends to become weaker not to cut the RG flow of $t_y$ too early. 
However, in above comparison  of the Peierls gaps, again, we assume $t_y/\epsilon_F$ to be small and to be of equal size for both $V_K$'s.
Therefore, our estimate of the Peierls energies is a conservative one and gives us actually a {\it lower bound on the energy gain}
coming from the Peierls gap formed by $V_{8k_F}$ compared to the one formed by  $V_{4k_F}$ in view of the renormalization of $t_y$ due to electron interactions.

In principal, there 
should be an optimal value of $V_K$ that via the RG cut-off would also determine the renormalized size of the Peierls gap. 
However,  there could be additional changes in energy such as, for example, electrostatic energy,
which are hard to quantify.
Given all these complications, we refrain  from attempting to find an explicit self-consistency equation here because, at best,
it only can give  a qualitative but not  quantitative answer.  Still, we can expect that there is an 
optimal finite $V_K$ based on the simple observation that there are two extreme cases: too weak and too strong potential $V_K$.

To summarize, we compare Peierls energies under the assumption that $t_y$ is small. We stress again that the Peierls gap in this 
regime  essentially does not depend on the potential strength $V_K$ (see Fig. 9) and is  determined by the hopping amplitude $t_y$,
see Eqs. (\ref{g_4}) - (\ref{g_4-8}).
Of course, $t_y$, in principle, could be different due to electrostatic effects for the $4k_F$ and the $8k_F$ case. However, 
it is quite plausible to assume that they should not differ substantially in strength  because all corrections to $t_y$ come mostly from  local reconfiguration of charges, 
so it is a perturbation to the initial value of $t_y$ determined for the uniform case ({\it i.e.}, $V_K=0$). 
In addition,  $t_y$ gets
renormalized by electron-electron interactions eventually, so the `initial' unrenormalized value of $t_y$ is not so important anyway. Here, we also refer 
back to the argument that the RG flow stops earlier in the $4k_F$ case than in the $8k_F$ case, strengthening our 
assumption of the $8k_F$ periodicity being dominant over $4k_F$. 

Given this mean-field picture, we  exclude  from now on the $4k_F$ case from further considerations and 
assume that the $8k_F$ contribution provides the dominant
 Fourier component (of the interaction-induced CDW) that leads to gaps.


\begin{thebibliography}{100}
\bibitem{Kane_lines_footnote} The anisotropic model  considered in Ref. \cite{kane_PRL,Kane_lines} has some partial similarity to the stripe model considered here, in that it also considers periodic arrangements of Luttinger liquid wires. However, in contrast to our model, the wires are arranged parallel to the edge of the strip and without supercells, and thus there is no resonant scattering condition of the type considered here. Moreover, the wires are fixed in position\cite{kane_PRL,Kane_lines} which gives rise
to a fixed lattice constant that is independent of the Fermi wavevector. As a result, the hierarchies of the FQHE states for fermions are different in the two models. In our case, we obtain all  QHE states of the Haldane-Halperin hierarchy and only those.
\end{thebibliography}
\end{document}